\documentclass[apjl,a4paper,12pt,useAMS]{emulateapj}
\usepackage{txfonts}
\usepackage{amssymb}

\usepackage{graphicx,graphics}
\usepackage{natbib}

\begin{document}
\title{Determining the luminosity function of {\it Swift} long gamma-ray bursts with pseudo-redshifts}
\author{Wei-Wei Tan\altaffilmark{1}}
\author{Xiao-Feng Cao\altaffilmark{2}}{}
\author{Yun-Wei Yu\altaffilmark{1}}

\altaffiltext{1}{Institute of Astrophysics, Central China Normal
University, Wuhan 430079, China {yuyw@phy.ccnu.edu.cn}}
\altaffiltext{2}{School of Physics, Huazhong University of Science
and Technology, Wuhan 430074, China}

\begin{abstract}
The determination of the luminosity function (LF) of gamma-ray
bursts (GRBs) is an important role for the cosmological applications
of the GRBs, which, however, is seriously hindered by some selection
effects due to redshift measurements. In order to avoid these
selection effects, we suggest calculating pseudo-redshifts for {\it
Swift} GRBs according to the empirical $L$-$E_p$ relationship. Here,
such a $L$-$E_p$ relationship is determined by reconciling the
distributions of pseudo- and real redshifts of redshift-known GRBs.
The values of $E_p$ taken from Butler's GRB catalog are estimated
with Bayesian statistics rather than observed. Using the GRB sample
with pseudo-redshifts of a relatively large number, we fit the
redshift-resolved luminosity distributions of the GRBs with a
broken-power-law LF. The fitting results suggest that the LF could
evolve with redshift by a redshift-dependent break luminosity, e.g.,
$L_b=1.2\times10^{51}(1+z)^2\rm erg~s^{-1}$. The low- and
high-luminosity indices are constrained to $0.8$ and $2.0$,
respectively. It is found that the proportional coefficient between
the GRB event rate and the star formation rate should
correspondingly decrease with increasing redshifts.
\end{abstract}

\keywords{gamma-ray burst: general}

\slugcomment{2013, APJL, ??, ??}

\section{Introduction}

Gamma-ray bursts (GRBs) are the most violent explosions in the
universe. Thanks to the {\it Swift} spacecraft, the number of GRBs
with measured redshifts has grown rapidly in the past decade.
Roughly speaking, redshifts have been measured for about one-third
of the total {\it Swift} GRBs. The highest redshift is reported to
be $z\sim9.4$ (Cucchiara et al. 2011). Some theoretical models even
predict that much more distant GRBs up to $z\sim 20$ could be
detected in the future (Band 2003; Bromm \& Loeb 2006; de Souza et
al. 2011). One of the most important astrophysical consequences of
the accumulated redshift data is the possible determination of the
GRB luminosity function (LF; Natarajan et al. 2005; Daigne et al.
2006; Salvaterra \& Chincarini 2007; Salvaterra et al. 2009, 2012;
Campisi et al. 2010; Wanderman \& Piran 2010; Cao et al. 2011). The
LF has a crucial role in cosmological applications of GRBs.

Nevertheless, strictly speaking, the present number of GRB redshifts
is still insufficiently large for a precise constraint on the GRB
LF, in particular, determining whether or not the LF evolves with
redshift. Meanwhile, the observational number distributions of GRB
redshifts and luminosities can be seriously distorted by some
unclear selection effects arising from redshift measurements (Cao et
al. 2011; Coward et al. 2012). On the one hand, the redshift
selection effects (RSEs) could be correlated to the optical
afterglow behaviors, the extinction of host galaxy (Jakobsson et al.
2004; Levan et al. 2006), and the redshift desert (Steidel et al.
2005). On the other hand, the RSEs can also be caused by
instruments, because the redshift measurements strongly depend on
the limiting sensitivity and spectral coverage of the spectroscopic
system (Greiner et al. 2008). Therefore, it is impossible to
describe the RSEs precisely from theoretical views, but some
effective parameterized expressions for the RSEs could be obtained
by carefully fitting the observational luminosity-redshift
distributions of GRBs (Cao et al. 2011).

As an alternative way to avoid the RSEs, one can convert some other
observational quantities of GRBs (e.g., the spectral peak energy,
the spectral indices, the variability indices, the afterglow related
quantities, etc) to a pseudo-redshift according to some
luminosity-indictor relationships. Such redshift estimating methods
were actually extensively discussed in the {\it Compton} BATSE era,
in particular, for determining the GRB formation history (e.g.,
Lloyd-Ronning et al. 2002; Murakami et al. 2003; Yonetoku et al.
2004). The tighter the relationship used, the closer the
pseudo-redshifts to the real ones. Since the calculation of
pseudo-redshifts is completely independent of the realistic redshift
measurements, the distortion of the distributions of redshifts and
luminosities by the RSEs can disappear in the pseudo-redshift GRB
sample. In addition, the number of the pseudo-redshifts, which could
be several times larger than the observed one, could facilitate a
more detailed statistics of {\it Swift} GRBs. Regardless, all of the
advantages of pseudo-redshifts are based on the precondition that
the used correlation is tight enough.

Until now, many empirical correlations between different properties
of GRBs have been proposed in the literature. The most popular ones
usually involve the spectral peak energy $E_p$ of GRBs. For example,
Amati et al. (2002) revealed that $E_p$ correlates with the
isotropically-equivalent released energy $E_{\rm iso}$, while
Ghirlanda et al. (2004) replaced $E_{\rm iso}$ by the
collimation-corrected energy $E_{\gamma}$. Liang \& Zhang (2005)
further suggested a triple correlation among $E_{p}$, the
isotropically-equivalent $\gamma$-luminosity, and the break time of
afterglow light curve. In this Letter, the GRB LF is of main
interest, so we will focus on the correlation between $E_{p}$ and
the peak luminosity $L$, using which Yonetoku et al. (2004) have
provided pseudo-redshifts for 689 BATSE GRBs and constrained their
LF.

In the next section, we test and adjust the $L$-$E_p$ relationship
by reconciling the distributions of pseudo- and real redshifts of
redshift-known {\it Swift} GRBs.
In Section 3, we constrain the LF parameters by fitting the
luminosity distributions of GRBs with pseudo-redshifts, where a
possible evolution of the LF is suggested. A summary and discussions
are provided in Section 4.

\section{$L$-$E_p$ relationship and pseudo-redshifts}
\begin{figure}
\centering\resizebox{0.5\textwidth}{!}{\includegraphics{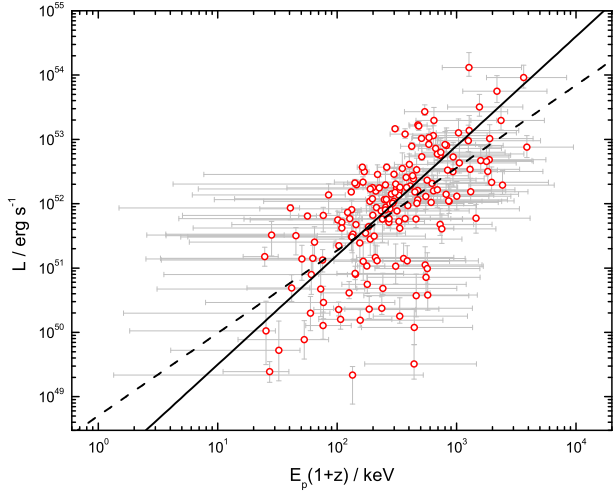}}
\caption{$L$-$E_p$ relationship of the 172 redshift-known {\it
Swift} GRBs (open circles). The dashed line represents the
least-squares fit, while the solid line is obtained by reconciling
the distributions of pseudo- and real redshifts.}\label{L-Ep}
\end{figure}
\begin{figure}
\centering\resizebox{0.5\textwidth}{!}{\includegraphics{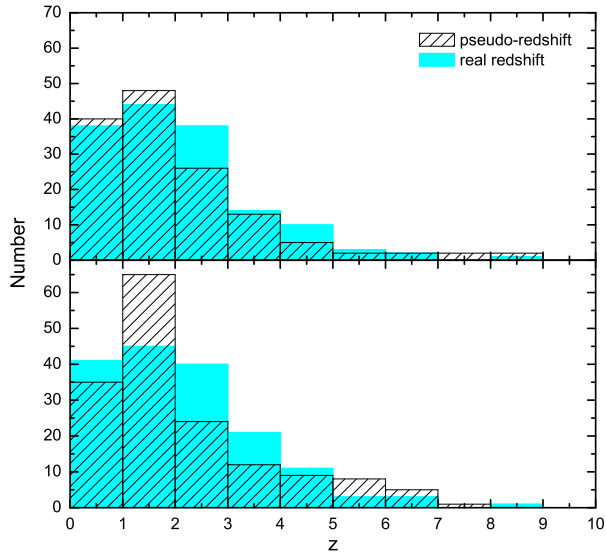}}
\caption{Comparisons between the number distributions of pseudo- and
real redshifts of the redshift-known GRBs. The top panel shows the
case where two distributions are closest to each other, while the
bottom panel is obtained with the least-squares fit to the $L$-$E_p$
relationship.}\label{n-n}
\end{figure}
Until GRB 120811C, in total there were 580 long-duration ($T_{90}>2$
s) GRBs detected by {\it Swift}, where 172 GRBs have been measured
at redshift.\footnote{These numbers are counted with the GRB catalog
provided by N. Butler; see http://butler.lab.asu.edu/{\it
Swift}/bat\_spec\_table.html (Butler et al. 2007, 2010).} Throughout
this Letter, only long-duration GRBs are considered. Additionally,
three GRBs with a luminosity $L<10^{49}\rm erg~s^{-1}$ have been
excluded, because they could belong to a distinct population called
low-luminosity GRBs (Soderberg et al. 2004; Liang et al. 2007). For
a redshift-known GRB, its luminosity can be calculated by $L=4\pi
d_l(z)^2P k(z)$, where $d_l(z)$ is the luminosity distance, $P$ is
the observed peak flux in the Burst Alert Telescope (BAT) energy
band 15-150 keV, and $k(z)\equiv \int_{1\rm keV}^{10^{4}\rm
keV}S(E')E'dE'/\int_{15(1+z)\rm keV}^{150(1+z)\rm keV}S(E')E'dE'$
(the primes represent rest-frame energy) converts the observed flux
into the bolometric flux in the rest-frame 1-$10^4$ keV. As is
widely accepted, the observed photon number spectrum $S(E)$ can be
well expressed by the empirical Band function (Band et al. 1993).
Here we take the related data including redshifts, peak fluxes, and
spectral peak energies from Butler's GRB catalog. It should be noted
that the peak energies are actually estimated by Bayesian statistics
but not directly observed, because the {\it Swift} BAT energy
bandpass is too narrow.
In Figure 1, we plot the 172 redshift-known GRBs in the $L$-$E_p$
plane, where a correlation between $L$ and $E_p$ appears. Such a
correlation was first proposed by Wei \& Gao (2003), Liang et al.
(2004), and Yonetoku et al. (2004) independently. The least-squares
fit to the $L$-$E_p$ correlation is shown by the dashed line in
Figure \ref{L-Ep}, which reads
\begin{eqnarray}
{L\over 10^{52}{\rm erg~s^{-1}}}=A\left[{E_p(1+z)\over 1 \rm
MeV}\right]^{\eta}\label{Ep-Lp},
\end{eqnarray}
with $A=3.47$ and $\eta=1.28$. According to the above equation, a
pseudo-redshift in principle can be derived from a pre-obtained peak
energy of a {\it Swift} GRB.
However, in view of the actual dispersion of the $L$-$E_p$
relationship, the pseudo-redshift is not expected to precisely equal
to the observationally measured one.

Therefore, for a less strict but statistically sound consideration,
here we suggest a new criterion for determining the $L$-$E_p$
relationship and calculating pseudo-redshifts. Instead of finding
sufficiently precise pseudo-redshifts for individual GRBs, we regard
all of the redshift-known GRBs as an entire statistical entity and
focus on the distribution of pseudo-redshifts. To be specific, first
we loosen the $L$-$E_p$ relationship by freeing the parameters $A$
and $\eta$ in Equation (1). Then the process of the determination of
a pseudo-redshift becomes as follows: (1) to assign a value to the
parameters $A$ and $\eta$; (2) to calculate pseudo-redshifts for the
172 redshift-known GRBs from the pre-assumed $L$-$E_p$ relationship,
where the errors of $E_p$ are ignored; (3) to compare the
distributions of pseudo- and real redshifts of the 172
redshift-known GRBs with the $\chi^2$ test; and (4) to find the most
likely values of $A$ and $\eta$ by minimizing the value of $\chi^2$.
In other words, our purpose is to reduce the discrepancy between the
distributions of pseudo- and real redshifts as much as possible.
Consequently, we obtain $A=7.93$ and $\eta=1.70$, as shown by the
solid line in Figure 1. This is obviously different from the
least-squares fit. With such a modified $L$-$E_p$ relationship, we
finally calculate pseudo-redshifts for 150 redshift-known GRBs, but
can not for the remaining 22 ones because their fluxes are  too low
with respect to their spectral peak energies. Both of the
distributions of pseudo- and real redshifts of the 150
redshift-known GRBs are presented in the top panel of Figure 2,
while the result corresponding to the least-squares fit is shown in
the bottom panel for a comparison.

By extending the modified $L$-$E_p$ relationship to all {\it Swift}
GRBs, 498 GRBs can be assigned a pseudo-redshift, occupying a
fraction 86\% of total {\it Swift} GRBs. Moreover, there are 38 GRBs
predicted to be beyond the redshift $z\sim10$, even at a redshift of
a few tens. Optimistically, one may expect to use these GRBs to
explore the universe at the beginning of reionization era. Figure 3
shows the 498 GRBs in the $L$-$z$ plane with their error bars
reflecting the uncertainties of peak energies of Butler's catalog
with a $90\%$ confidence. Strictly speaking, the errors of a great
number of GRBs could be too large for an exact statistics, which
makes our following results subjecting to large uncertainty. Anyway,
the open circles in Figure 3 representing the most likely parameters
of the GRBs seem still to distribute ``normally" in the $L$-$z$
plane, by comparing with the $L$-$z$ distribution of redshift-known
GRBs (e.g., see Figure 1 in Cao et al. 2011). Therefore, as a
perilous attempt, in the following statistics we only take into
account the most likely parameters and ignore their error bars.

\begin{figure}
\centering\resizebox{0.5\textwidth}{!}{\includegraphics{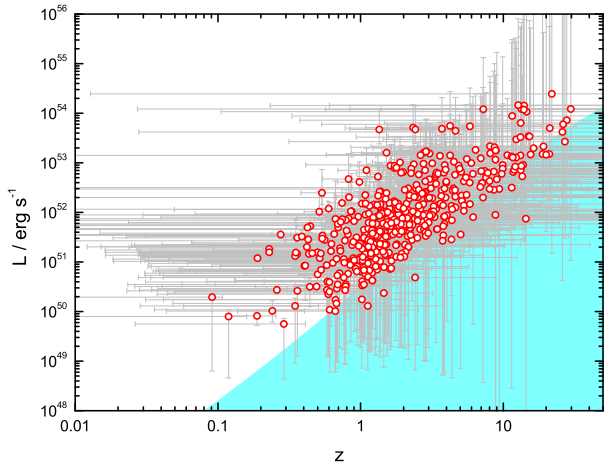}}
\caption{Luminosity-redshift distribution of 498 GRBs with
pseudo-redshifts, where the open circles represent the most likely
values and the error bars are determined by the uncertainties of
peak energies of Butler's catalog with a $90\%$ confidence. The
shaded region is determined by the lower cutoff luminosity (see
Equation (\ref{sencitivity})). }\label{L-z}
\end{figure}
\section{Luminosity function}
As the main interest of this Letter, we constrain the GRB LF with
the pseudo-redshift-obtained GRBs. In comparison with previous
works, this work could have two advantages: (1) the RSEs have been
removed, which significantly reduces the uncertainty of the model,
and (2) the GRB sample is enlarged by about three times, which makes
it possible to provide redshift-resolved luminosity distributions.
Our statistics would be restricted to below the redshift $3.5$,
because the star formation history above $z\sim3.5$ is unclear at
present (Hopkins \& Beacom 2006; Bouwens et al. 2007, 2011; Oesch et
al. 2010; Yu et al. 2012; Coe et al. 2012; Tan \& Yu 2013). For
relatively low redshifts, the star formation history can be
described by (Hopkins \& Beacom 2006),
\begin{eqnarray}
\dot{\rho}_\star(z)\propto \left\{~~\begin{array}{ll}(1+z)^{3.44},&
~~~~z\leq0.97,\,\\
(1+z)^{0}, & ~~~~0.97<z\leq3.5,\,
\end{array}\right.\label{star formation rate}
\end{eqnarray}
with the local star formation rate $\dot{\rho}_\star(0)=0.02 ~
M_{\odot}\rm yr^{-1} Mpc^{-3}$.

It is natural to consider that the paucity or absence of GRBs in the
range of relatively low luminosities as shown by the shaded region
in Figure 3 is caused by the multiple thresholds of all related
telescopes, especially the {\it Swift} BAT. However, because of the
very complicated trigger processes of the BAT, it seems impossible
to exactly express the BAT threshold. We only know that the trigger
probability of the BAT could increase with increasing gamma-ray
brightness and eventually approach one. Therefore, in order to avoid
the uncertainty arising from the BAT trigger probability, we take a
sufficiently high flux $P_{\rm lc}=2\times 10^{-8}~\rm
erg~s^{-1}~cm^{-2}$ as the lower cutoff of peak flux for our
statistics. Above $P_{\rm lc}$, the trigger probability can
basically be considered to unity. The corresponding lower cutoff
luminosity, which is shown by the upper boundary of the shaded
region in Figure 3, can be calculated by
\begin{eqnarray}
L_{\rm lc}(z)=4\pi d_L(z)^2 P_{\rm lc} k(z),\label{sencitivity}
\end{eqnarray}
where the $k$-correction factor is calculated with a typical
(averaged) rest-frame peak energy $E'_p\sim 150~\rm keV$. As a
result, 17 GRBs with $L<L_{\rm lc}$ are excluded and 361 GRBs
remain.

\begin{figure}\centering
\centering\resizebox{0.4\textwidth}{!}{\includegraphics{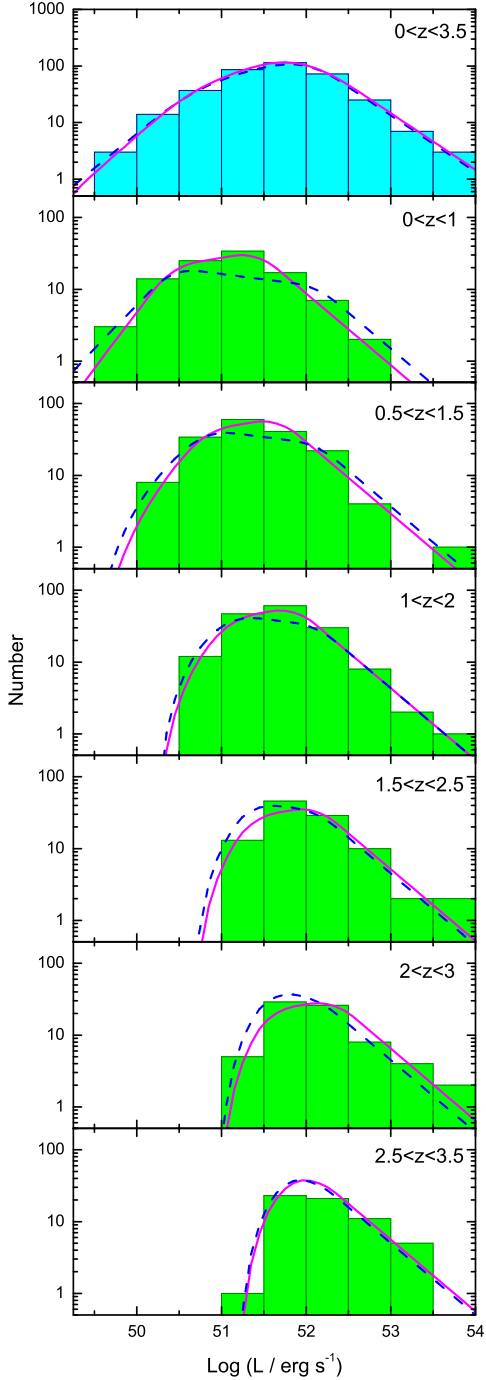}}
\caption{Distributions of GRB numbers in different luminosity bins
for different redshift intervals as labeled. Fittings to the
luminosity distributions with a constant and an evolving LF are
presented by the dashed and solid lines, respectively. Note that the
GRB numbers presented here are actually counted with large
uncertainties of the luminosities, as indicated in Figure 3, which
are, however, not displayed for simplicity.}\label{N-L}
\end{figure}

\begin{figure}
\centering\resizebox{0.4\textwidth}{!}{\includegraphics{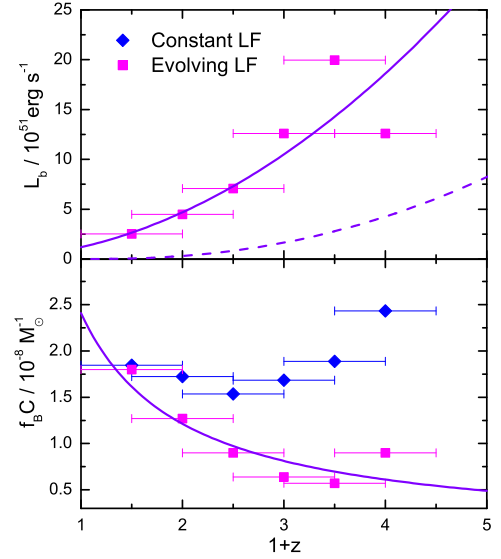}}
\caption{Redshift-dependences of parameters $L_b$ and $f_BC$ in both
constant and evolving LF cases, where the horizon error bars
represent the width of the redshift intervals. The errors of the
parameter values could be very large but not displayed due to the
difficulty of the error estimation. In the evolving LF case, two
empirical power-law fittings are provided by the solid lines. The
adopted lower cutoff luminosity is presented by the dashed line in
the top panel for a comparison. }\label{Lb-C}
\end{figure}
In view of the relatively large number of the remaining GRBs, we
divide the adopted GRB sample into six redshift intervals as
$0<z<1$, $0.5<z<1.5$, $1<z<2$, $1.5<z<2.5$, $2<z<3$, and
$2.5<z<3.5$. The adjacent intervals are taken to overlap with each
other just in order to obtain a sufficiently large GRB number for
each redshift interval. Then, from the second to seventh top panels
in Figure 4, we display the corresponding redshift-resolved
luminosity distributions independently, and meanwhile the combined
distribution of all redshift ranges is presented in the first top
panel. In the theoretical aspect, the GRB number within the
luminosity bin $L_1<L<L_2$ for a redshift interval $z_1<z<z_2$ can
be calculated by
\begin{eqnarray}
N&=&{\Delta\Omega\over 4\pi} T
\int^{z_2}_{z_1}\int^{L_2}_{\max[L_1,L_{\rm
lc}(z)]}\Phi_z(L)\dot{R}(z)dL{dV(z)\over 1+z}, \label{expected
number}
\end{eqnarray}
where $(\Delta\Omega/4\pi)\sim0.1$ is the field view of the BAT,
$T\sim7 \rm yr$ the observational period, and $dV(z)$ the comoving
volume. The observational GRB production rate can be connected to
star formation rate as
\begin{eqnarray}
\dot{R}(z)=f_BC ~\dot{\rho}_\star(z),\label{GRB rate}
\end{eqnarray}
where $f_B$ is the beaming degree of GRB outflows and the
proportional coefficient $C$ could arise from the particularities of
GRB progenitors (e.g., mass, metallicity, magnetic field, etc). For
the LF $\Phi_z(L)$, two popular competitive forms, including the
broken-power law and the single-power law with an exponential cutoff
at low luminosity, are tested in Cao et al. (2011). As a result, the
broken-power-law LF as
\begin{eqnarray}
\Phi_z(L)\propto\left\{~\begin{array}{ll}\left({L\over
L_b(z)}\right)^{-v_1},~~~~& L\leq L_b(z),\,\\
\left({L\over
L_b(z)}\right)^{-v_2},~~~~&L>L_{b}(z),\,\end{array}\right.
\label{luminosity function}
\end{eqnarray}
is suggested to be more consistent with observations. The
normalization coefficient of the LF is taken 
with an assumed minimum luminosity $L_{\min}=10^{49}\rm
~erg~s^{-1}$. To summarize, for a fitting to a GRB luminosity
distribution, we need to determine values for four model parameters
as $f_BC$, $L_b$, $\nu_1$, and $\nu_2$.

For a general consideration, here we fit the GRB luminosity
distributions in two different ways, i.e., with a constant and an
evolving LF, which are presented by the dashed and solid lines in
Figure 4, respectively. In both cases, the values of $\nu_1$ and
$\nu_2$ are considered to be constant. As suggested in Cao et al.
(2011), the value of the high-luminosity index $\nu_2$ could be
found directly from the distribution of high-luminosity GRBs,
because the telescope thresholds nearly cannot affect the detection
of these GRBs. Then we get $\nu_2=2.0$, which is the same as that in
Cao et al. (2011). Subsequently, (1) with a constant LF assumption,
we constrain the model parameters by fitting the combined luminosity
distribution of all GRBs, which gives rise to the best-fitting
parameters $L_b=1.3\times10^{52}$ and $\nu_1=1.2$.\footnote{The
lower value of $\nu_1$ than that in Cao et al. (2011) may indicate
that the RSEs are overestimated in Cao et al. (2011).} With the
determined $L_p$, $\nu_1$, and $\nu_2$, we further fit the six
redshift-resolved luminosity distributions to find the best-fitting
values of $f_BC$ in different redshift ranges. The results are
presented in the bottom panel of Figure (diamonds), where the last
data at $1+z=4$ lead us to see an increase of $f_BC$ with increasing
redshift. This is qualitatively consistent with previous findings of
the evolution effect (e.g., Kistler et al. 2009). However, if we
remove the last data, the redshift-dependence of $f_BC$ could become
ambiguous or, at least, very weak. (2) In the evolving LF model, we
should fit the six redshift-resolved luminosity distributions
independently rather than with a predetermined $L_p$. The
best-fitting values of $L_p$ and $f_BC$ for different redshifts are
also presented in Figure 5 (squares). The constant low-luminosity
index is simultaneously determined to $\nu_1=0.8$. As shown by the
solid lines in Figure 5, on the one hand, an obvious evolution of
the break luminosity $L_b$ appears as
\begin{eqnarray}
L_b&=&1.2\times10^{51} (1+z)^{2}~\rm erg~s^{-1},
\end{eqnarray}
which is qualitatively consistent with the result in Yonetoku et al.
(2004) for BATSE GRBs. Moreover, the difference between the redshift
dependences of $L_b(z)$ and $L_{\rm lc} (z)$ suggests that the
evolution of the LF is probably intrinsic rather than observational.
Such an LF evolution indicates that higher-redshift GRBs could be
much brighter than the lower-redshift ones. On the other hand, the
coefficient $f_BC$ is found to decrease with increasing redshift as
\begin{eqnarray}
f_BC&=&2.4\times10^{-8} (1+z)^{-1} M_{\odot}^{-1},
\end{eqnarray}
which is completely opposite to that of the previous understanding
with a constant LF.

Finally, Figure 4 shows that both the constant and evolving LF
models can provide a perfect fitting to the combined luminosity
distribution of all GRBs. In other words, it is impossible to
distinguish the two models by the combined distribution. However,
for the six redshift-resolved luminosity distributions, it is
clearly shown that the fittings with an evolving LF are always
better than the ones with a constant LF, in particular, for
relatively low redshifts. Therefore, we prefer to conclude that an
evolving LF is more favored by the pseudo-redshift GRB sample.

\section{Summary and discussions}
In view of their association with Type Ib/c supernovae and the
bright gamma-ray emission (Stanek et al. 2003; Hjorth et al. 2003;
Chornock et al. 2010), GRBs are usually suggested to trace the
cosmic star formation history. However, due to the thresholds of
telescopes, the conversion from the GRB event rate to star formation
rate is strongly dependent on the form of the LF, and moreover the
determination of the LF is seriously hindered by the RSEs. One
viable method is to take only the high-luminosity GRBs into account,
as was done in Kistler et al. (2009) and Wang \& Dai (2011). Such a
method could become invalid if the LF is redshift-dependent.
Alternatively, in this Letter, we suggest to use an empirical GRB
relationship (i.e., $L$-$E_p$ relationship) to calculate
pseudo-redshifts for {\it Swift} GRBs, so that the RSEs can be
avoided in the new redshift sample. In view of the insufficient
tightness of the adopted $L$-$E_p$ relationship, we replace the
least-squares fit to the relationship by a modified one with which
the distribution of the pseudo-redshifts can be closest to the
observational one. Consequently, a GRB sample of a large number is
obtained, which makes it possible to analyze the GRB luminosity
distributions in different redshift ranges. As found by Yonetoku et
al. (2004) for BATSE GRBs, here we also find the LF of {\it Swift}
GRBs could evolve with redshift by a redshift-dependent break
luminosity $L_b$. Such an evolving LF also changes our understanding
of the intrinsic connection between the GRBs and stars, i.e., the
parameter $f_BC$ should decrease (but not increase as previously
considered) with increasing redshifts. In other words, both the GRB
production efficiency and the luminosities of the produced GRBs
should be very different at different cosmic times, which may
provide some new constraints on the properties of GRB progenitors
and central engines.

\acknowledgements This work is supported by the National Natural
Science Foundation of China (Grant No. 11103004) and the Funding for
the authors of National Excellent Doctoral Dissertations of China
(Grant No. 201225).


\begin{thebibliography}{99}


\bibitem{Amati 02}Amati, L., Frontera, F., Tavani, M., et al. 2002,
A\&A, 390, 81

\bibitem{Band 93} Band, D., Matteson, J. Ford, L. et al. 1993, ApJ, 413, 281

\bibitem{Band 03}Band, D. L. 2003, ApJ, 588, 945

\bibitem{Bouwens 07}Bouwens, R. J., Illingworth, G. D., Franx, M., \& Ford, H. 2007, ApJ,
670, 928
\bibitem{Bouwens 11}Bouwens, R. J., Illingworth, G. D., Labbe, I., et al. 2011,
Natur, 469, 504

\bibitem{Bromm 06}Bromm, V., \& Loeb, A. 2006, ApJ, 642, 382

\bibitem{Butler 10}Butler, N. R., Bloom, J. S., \& Poznanski, D. 2010, ApJ, 711,
495

\bibitem{Butler 07}Butler, N. R., Kocevski, D., Bloom, J. S., \& Curtis, J. L. 2007, ApJ,
671, 656

\bibitem{Campisi et al. 10}Campisi, M. A., Li, L.-X., \& Jakobsson, P. 2010, MNRAS,407, 1972

\bibitem{Cao 11}Cao, X. F., Yu, Y. W., Cheng, K. S., \& Zheng, X. P. 2011, MNRAS, 416, 2174

\bibitem{Chornock 03}Chornock, R., Berger, E., Levesque, E. M., et
al. 2010, ApJL, submitted (arXiv:1004.2262)

\bibitem{Coe 12}Coe, D., Umetsu, K., Zitrin, A., et al. 2012, ApJ, 757, 22

\bibitem{Coward 12}Coward, D.M., Howell1, E. J., Branchesiet, M., et al. 2012, arXiv: 1210.2488

\bibitem{Cucchiara 11}Cucchiara, A., Levan, A. J., Fox, D. B., et
al. 2011, ApJ, 736, 7

\bibitem{Daigne et al. 06}Daigne, F., Rossi, E. M., \& Mochkovitch, R. 2006, MNRAS, 372, 1034

\bibitem{de Souza 11}de Souza, R. S., Yoshida, N., \& Ioka, K. 2011, A\&A,
533, 32


\bibitem{Ghirlanda 04}Ghirlanda, G., Ghisellini, G., \& Lazzati, D.
2004, ApJ, 616, 331


\bibitem{Greiner 08}Greiner J., Bornemann, W., Clemens, C., et al. 2008, PASP, 120, 405


2004, ApJ, 615, L73


\bibitem{Hjorth 03}Hjorth, J., Sollerman, J., M{\o}ller, P., et al.
2003, Natur, 423, 847

\bibitem{Hopkins 06}Hopkins, A. M., \& Beacom, J. F. 2006, ApJ,
651, 142

\bibitem{Jakobsson 04}Jakobsson, P., Hjorth, J., Fynbo, J. P. U., et al. 2004, A\&A, 427, 785

\bibitem{Kistler 09}Kistler, M. D., Y$\rm \ddot{u}$ksel, H., Beacom, J. F., et al. 2009, ApJL, 705, L104

\bibitem{Levan 06}Levan, A., Fruchter, A., Rhoads, J., et al. 2006, ApJ, 647, 471

\bibitem{}Liang, E. W., Dai, Z. G., \& Wu, X. F. 2004, ApJL, 606, L29

\bibitem{Liang 05}Liang, E. W., \& Zhang, B. 2005, ApJ, 633, 611

\bibitem{Liang 07}Liang, E. W., Zhang, B., Virgili, F., \& Dai, Z.
G. 2007, ApJ, 662, 1111

\bibitem{Lloyd-Ronning 02}Lloyd-Ronning, N. M., Fryer, C. L., \& Ramirez-Ruiz,
E. 2002, ApJ, 574, 554

\bibitem{Murakami 03}Murakami, T., Yonetoku, D., Izawa, H., \& Ioka,
K. 2003, PASJ, 55, L65

\bibitem{Natarajan et al. 05} Natarajan, P., Albanna, B., Hjorth, J., et al. 2005, MNRAS, 364, L8

\bibitem{}Oesch, P. A., Bouwens, R., J., Illingworth, G. D., et al. 2010, ApJL, 709, L16


\bibitem{Salvaterra 07}Salvaterra, R., \& Chincarini, G. 2007, ApJL, 656, L49

\bibitem{Salvaterra et al. 09}Salvaterra, R., Guidorzi, C., Campana, S., Chincarini, G., \& Tagliaferri, G. 2009, MNRAS, 396, 299

\bibitem{Salvaterra et al. 12}Salvaterra, R., Campana, S., Vergani, S.
D., et al. 2012, ApJ, 749, 68



\bibitem{Soderberg 04}Soderberg, A. M., Kulkarni, S. R., Berger, E.,
et al. 2004, Natur, 430, 648


\bibitem{Stanek 03}Stanek, K. Z., Matheson, T., Garnavich, P. M., et
al. 2003, ApJH, 591, L17

\bibitem{Steidel 05}Steidel, C., Shapley, A., Pettini, M., et al. 2005, in Proc. ESO Workshop
in Venice, Italy, 13-16 October 2003, Multiwavelength Mapping of
Galaxy Formation and Evolution, ed. A. Renzini and R. Bender
(Berlin: Springer) 169

\bibitem{Tan 13}Tan, W. W., Yu, Y. W. 2013, Sci. China Phys. Mech. Astron.,
56, 1029

\bibitem{Wanderman 10}Wanderman, D., \& Piran, T. 2010, MNRAS, 406,
1944

\bibitem{Wang 11}Wang, F. Y., \& Dai, Z. G. 2011, ApJL, 727, L34

\bibitem{}Wei, D. M., \& Gao, W. H. 2003, MNRAS, 345, 743

\bibitem{Yonetoku 04} Yonetoku, D., Murakami, T., Nakamura, T., et
al. 2004, ApJ, 609, 935

\bibitem{Yu 12}Yu, Y. W., Cheng, K. S., Chu, M. C., \& Yeung, S. 2012,
JCAP, 07, 023



\end{thebibliography}
\end{document}